\documentclass[12pt,letterpaper]{article}

\usepackage{amsmath,amssymb,calc}
\usepackage{graphicx}
\usepackage{color}

\usepackage{cite}

\newcommand\be{\begin{equation}}
\newcommand\ee{\end{equation}}
\newcommand{\bea}{\begin{eqnarray}}
\newcommand{\eea}{\end{eqnarray}}

\newcommand{\nn}{\nonumber}
\newcommand{\pd}{\partial}

\def\id{\protect{{1 \kern-.28em {\rm l}}}}

\def\id{\protect{{1 \kern-.28em {\rm l}}}}

\begin{document}

\begin{titlepage}
\begin{center}
\hfill \\
\vspace{1cm}
{\Large {\bf New Effects in Multifield Cosmology\\ Coupled to the Gauss-Bonnet Term\\[3mm] }}

\vskip 1.5cm
{\bf Lilia Anguelova${}^a$\footnote{anguelova@inrne.bas.bg}, Sonia Paban${}^b$\footnote{sppaban@fas.harvard.edu}\\
\vskip 0.5cm  {\it ${}^a$ Institute for Nuclear Research and Nuclear Energy,
Bulgarian Academy of Sciences, Sofia 1784, Bulgaria\\
${}^b$ Leinweber Institute for Theoretical Physics and Department of Physics, Harvard University,
Cambridge, MA, 02138, USA }}

\vskip 6mm

\end{center}

\vskip .1in
\vspace{0.5cm}

\begin{center} {\bf Abstract}\end{center}

\vspace{-1cm}

\begin{quotation}\noindent

We study the rapid-turn regime of multifield cosmology coupled to the Gauss-Bonnet term. We derive the corresponding modified relations between the inflationary Hubble and potential slow-roll parameters, in the two-field case. These relations show that including the Gauss-Bonnet coupling broadens the possibilities for realizing rapid turn inflation and for satisfying the swampland conjectures. We also derive a consistency condition relating the scalar field-space metric to the scalar potential and the Gauss-Bonnet coupling, which two-field slow-roll and rapid-turn inflationary solutions have to satisfy. Finally, we find general conditions under which phantom-like behavior is possible in multifield dark energy models coupled to the Gauss-Bonnet term. Notably, although we do not consider a direct interaction between dark matter and dark energy, dark matter is crucial for the occurrence of a phantom regime.

\end{quotation}

\end{titlepage}

\eject

\tableofcontents

\section{Introduction}
\setcounter{equation}{0}

A plethora of precise observational data accumulated over recent decades has transformed modern cosmology. The parameters of the standard cosmological model, known as $\Lambda$CDM, are now established with considerable precision, providing a comprehensive understanding of the Universe’s composition and evolution. Despite its many successes, however, $\Lambda$CDM remains a phenomenological model that relies on components—dark matter and dark energy—whose fundamental nature is still unknown. Moreover, several cosmological tensions have emerged in recent years, reflecting discrepancies among estimates of the same cosmological parameter derived from different datasets. Together, these developments strongly motivate investigating theories beyond the standard cosmological model.

In the search for a more fundamental understanding consistent with quantum gravity, multifield cosmology—arising from the coupling of multiple scalar fields to Einstein gravity—provides a natural framework \cite{OOSV,GK,OPSV,AP,BPR,Baumann:2014nda}. Furthermore, it yields novel effects compared with those in the better-studied single-scalar models of cosmic acceleration (for an incomplete list, see \cite{Malik:2008im, Chen:2009zp, Baumann:2014nda, Renaux-Petel:2015mga, Achucarro:2017ing, Brown:2017osf, BM, TB, Fumagalli:2019noh, Achucarro:2022qrl} ). In multifield models of inflation, for example, slow-roll can occur on steep potentials \cite{HP, AP, Brown:2017osf, Aragam:2019omo, ACPRZ, Freigang:2023ogu}, and primordial black holes can be easily generated \cite{PSZ,FRPRW,LA1_pbh,LA2_pbh}. Multifield models of dark energy can also produce observational signatures that distinguish them from a cosmological constant, even when the equation-of-state parameter is close to $-1$ \cite{ASSV,EASV,ADGW,ADGW2}. These possibilities have attracted growing interest in multifield cosmology. However, existing studies have largely set aside another natural ingredient of the low-energy effective actions of fundamental gravitational theories: higher-curvature terms. Here, we begin to investigate how incorporating these terms can modify the effects described above and give rise to new phenomena.\footnote{Although higher curvature terms are generically suppressed by powers of the ratio $H/\Lambda_s$, where $H$ is the Hubble parameter and $\Lambda_s$ is the species scale \cite{vandeHeisteeg:2023ubh}, their contributions need not be negligible compared to the slow-roll parameters.} 

The richer phenomenology of multifield models, relative to that of single-scalar ones, arises from the so-called rapid-turn regime. This regime is characterized by solutions to the background field equations whose trajectories in field space are strongly non-geodesic. The deviation of a trajectory from a geodesic is quantified by its turning rate. We will examine how higher-curvature terms modify key relations among important characteristics of solutions with large turning rates. This analysis yields broader possibilities for realizing the rapid-turn regime, as well as new features of these cosmological models that are of observational interest.

As an initial step toward a more systematic understanding of the role of higher-curvature terms in the multifield context, we focus here on the Gauss-Bonnet term. Our motivation stems both from its appearance in string-theoretic effective actions \cite{BZ,GS,GMV} and from its relative technical simplicity. In four dimensions, the Gauss-Bonnet term is a total derivative and therefore does not contribute to the dynamics on its own. However, when coupled to scalar fields, it affects their equations of motion through derivatives of the coupling function. Models of slow-roll inflation incorporating this higher-curvature term have already been studied in the single-scalar context; see, for instance, \cite{GuS,JHG,KLLT}. More recently, \cite{SK} computed the quadratic effective action for inflationary perturbations in two-field models coupled to the Gauss-Bonnet term. However, the definition of the slow-roll approximation in that work was too restrictive, as will become clear shortly. 

We establish the proper generalization of the rapid turn regime, in two-field models, in the presence of the Gauss-Bonnet (GB) term and compute the modified relation between the first slow-roll parameter $\varepsilon$\,, characterizing the rate of spacetime expansion, and its counterpart $\varepsilon_V$\,, determined from the gradient of the scalar potential. The contribution of the GB term can enhance the difference between $\varepsilon$ and $\varepsilon_V$\,, that is due to rapid turning, or can even drive it. Thus, the coupling to GB opens new avenues for consistency with the swampland conjectures. We also show that certain Hessian component of the GB coupling function, that was assumed to be small during slow-roll in \cite{SK}, is in fact non-negligible not only in a rapid turn regime, but also in a slow-roll {\it slow-turn} one. Furthermore, we derive relations between the Hessian components of the potential and of the GB coupling function, that are relevant for the generalization of the second potential slow-roll parameter $\eta_V$ to the multifield context. These relations show that the GB contribution can affect the amount of turning, needed to resolve the $\eta$-problem of UV-completions of inflation. Finally, we derive a consistency condition that two-field slow-roll and rapid-turn inflationary trajectories in field space have to satisfy, which relates the GB coupling function to the scalar potential and field-space metric. This relation generalizes the consistency condition of \cite{LA,WIA}.

Higher curvature terms can enrich not only multifield inflationary models, but also models of dynamical dark energy (DE). This is particularly relevant in view of recent observational indications \cite{DESI,CCSGVM} that the DE equation-of-state parameter $w_{DE}$ may have been less than $-1$ in the not too distant past of the Universe. Such a `phantom' regime cannot be achieved neither in standard quintessence (i.e. single-scalar DE) nor in standard multifield cosmology. However, there are examples of phantom-like behavior in single-field models interacting with a GB term \cite{NOS,NOO}. Note that this is simply due to the coupling of the scalar to the higher curvature term; neither exotic matter, nor exotic forms of energy are involved. Here we derive general conditions for achieving $w_{DE} < -1$ in multifield models coupled to the Gauss-Bonnet term. Interestingly, although we do not consider direct interaction between DE and dark matter (DM), unlike in \cite{BOVW,AKW,BRWX}, the phantom-like behavior can only occur if DM is non-negligible, as is the case in the late Universe.

This paper is organized as follows. In Section \ref{EoMsFB}, focusing on two-field models, we briefly review the equations of motion and the kinematic field-space basis, which is necessary to define the turning rate. In Section \ref{SR-RT}, we introduce the relevant slow-roll and rapid-turn parameters and discuss the proper generalization of the rapid-turn regime. Section \ref{EpEpV} derives the GB-modified relation between the inflationary parameters $\varepsilon$ and $\varepsilon_V$\,, while Section \ref{HesVf} studies the Hessian components of the GB coupling function $f$\,. We derive their contributions to corresponding components of the Hessian of the scalar potential $V$ and discuss the implications of these results for the relation between the inflationary parameters $\eta$ and $\eta_V$\,. In Section \ref{ConsC-SRRT}, we obtain a relation between the field-space metric, scalar potential and the GB coupling function, which is a necessary condition for slow-roll and rapid-turn inflation in two-field models. Finally, in Section \ref{DEw} we find general conditions, under which a GB-coupled multifield dark energy model can exhibit phantom-like behavior. We conclude, in Section \ref{Disc}, with a brief summary and a discussion of open research directions.

\section{Equations of motion and field-space basis} \label{EoMsFB}
\setcounter{equation}{0}

We will study an extension of multifield cosmology, obtained by adding the Gauss-Bonnet term to the action for multiple scalars minimally coupled to Einstein gravity. Thus, we consider the action:
\be \label{Action}
S \,= \int \!d^4 x \sqrt{- \det g} \!\left[ \frac{R}{2} - \frac{1}{2} \,G_{IJ}(\phi) \,g^{\mu \nu} \,\pd_{\mu} \phi^I \,\pd_{\nu} \phi^J - V (\phi) + \frac{f(\phi)}{8} \,R^2_{GB} \right] \,\,,
\ee
where $R^2_{GB}$ is the standard Gauss-Bonnet term:
\be
R^2_{GB} = R^2 - 4 R_{\mu \nu} R^{\mu \nu} + R_{\mu \nu \rho \sigma} R^{\mu \nu \rho \sigma} \,\,\, ,
\ee
while $g_{\mu \nu}$ is the spacetime metric and $G_{IJ}$ is the curved metric on the field space parameterized by the scalars $\phi^I$\,. We work in units where $M_P=1$. We focus on the two-field case, namely the indices $I,J = 1,2$, in the following. We also treat the coupling function $f(\phi)$ and the potential $V(\phi)$ as arbitrary functions, modulo the assumption that the potential is positive-definite.

The standard Ansatze for an isotropic and homogeneous cosmological background are:
\be \label{Bkgr}
ds_g^2 = -dt^2 + a(t)^2 d\vec{x}^2 \quad , \quad \phi^I = \phi^I_0 (t) \quad .
\ee
Using these in (\ref{Action}) leads to the following equations of motion. For the background scalars one has:
\be \label{sEoM}
D_t \dot{\phi}^I_0 + 3 H \dot{\phi}_0^I + G^{IJ} V_J - 3 G^{IJ} f_J H^2 (H^2 + \dot{H}) = 0 \,\,\, ,
\ee
where
\be
\quad H \equiv \frac{\dot{a}}{a} \,\,\, , 
\ee
as usual, and
\be \label{Defs}
V_J \equiv \pd_{\phi_0^J} V \quad , \quad f_J \equiv \pd_{\phi_0^J} f \quad , \quad 
D_t \dot{\phi}^I_0 \equiv \dot{\phi}_0^J \,\nabla_J \dot{\phi}^I_0 = \ddot{\phi}^I_0 + \Gamma^I_{JK} \dot{\phi}_0^J \dot{\phi}^K_0 \,\,\, .
\ee
The field equations for the background space-time metric are:
\be \label{gEoM}
3H^2 = \frac{1}{2}G_{IJ} \dot{\phi}_0^I \dot{\phi}_0^J + V - \,3 f_I \dot{\phi}_0^I H^3 \,\,\, ,
\ee
where note that $\dot{f} = f_I \dot{\phi}_0^I$\,, and
\be \label{gEoM2}
- \frac{1}{2}G_{IJ} \dot{\phi}_0^I \dot{\phi}_0^J + V = 2 (1+H \dot{f}) (H^2 + \dot{H}) + H^2 (1+\ddot{f}) \,\,\, .
\ee
For future use, we also note that (\ref{gEoM2}) can be rewritten as follows:
\be \label{gEoM3}
- \frac{1}{2} G_{IJ} \dot{\phi}_0^I \dot{\phi}_0^J + V = 3H^2 + 2 \dot{H} + 2 \dot{f} H (H^2 + \dot{H}) + \ddot{f} H^2\, .
\ee
Substituting $V$ from (\ref{gEoM}) into (\ref{gEoM3}) gives:
\be \label{Hd}
2 \dot{H} = - G_{IJ} \dot{\phi}_0^I \dot{\phi}_0^J + \dot{f} H^3 - 2 \dot{f} H \dot{H} - \ddot{f} H^2 \,\,\, .
\ee

Note that, although (\ref{sEoM}) may suggest that it could be useful to introduce an 'effective potential' of the form $V_{eff} = V -3 f H^2 (H^2 + \dot{H}) $\,, from (\ref{gEoM}) it is clear that this is not the case. Indeed, in the following we will see that the potential $V$ plays a role independently of $f$. And, of course, unlike $V$\,, the coupling function $f$ enters all dynamical equations only via its derivatives, in accordance with the fact that the Gauss-Bonnet term is a topological invariant in four dimensions.

It is useful to introduce the following basis in field space:
\be \label{basis}
T^I = \frac{\dot{\phi}^I_0}{\dot{\sigma}} \quad , \quad N_I = (\det G)^{1/2} \epsilon_{IJ} T^J\quad ,\quad \dot{\sigma}^2 = G_{IJ} \dot{\phi}^I_0 \dot{\phi}^J_0 \,\,\, ,
\ee
where $\epsilon_{IJ}$ is the standard antisymmetric symbol with $\epsilon_{12} = - \epsilon_{21} = 1$\,. Note that $T^I$ is a tangent vector, while $N^I$ is a normal vector, with respect to the background trajectory $\left( \phi_0^1 (t) , \phi_0^2 (t) \right)$ in field space. In terms of this basis, one can define the turning rate of the trajectory:
\be
\Omega = - N_I D_t T^I \quad , \quad D_t T^I \equiv \dot{\phi}_0^J \,\nabla_J T^I \,\,\, ,
\ee
which is a quantity that measures its deviation from a field-space geodesic. For future use, note that:
\be \label{TN-Om}
D_t T^I = - \Omega N^I \qquad {\rm and} \qquad D_t N^I = \Omega T^I \quad .
\ee

In the following, we will also need the projections of the scalar field equations of motion (\ref{sEoM}) along the above basis. The $T^I$ projection gives:
\be \label{sigma-dd}
\ddot{\sigma} + 3 H \dot{\sigma} + V_T - 3 f_T H^4 (1 - \varepsilon) = 0 \,\,\, ,
\ee
where $V_T \equiv T^I V_I$ and $f_T \equiv T^I f_I$\,, while contracting (\ref{sEoM}) with $N_I$ leads to:
\be \label{Omf}
\Omega = \frac{V_N}{\dot{\sigma}} - \,\frac{3 f_N}{\dot{\sigma}} H^2 (H^2 + \dot{H}) \,\,\, ,
\ee
where $V_N \equiv N_I V^I$ and $f_N \equiv N_I f^I$.

\section{Slow-roll and rapid-turn parameters} \label{SR-RT}
\setcounter{equation}{0}

To characterize the physics of models of cosmic acceleration within this framework, it is convenient to introduce a set of parameters that describe the regimes of slow-roll and rapid turn spacetime expansion. In standard multifield cosmology with two scalars, these parameters are:
\be \label{Param_def}
\varepsilon = - \frac{\dot{H}}{H^2} \quad , \quad \eta_{\parallel} = - \frac{\ddot{\sigma}}{H \dot{\sigma}} \quad , \quad \eta_{\perp} = \frac{\Omega}{H} \,\,\, ,
\ee
where the notation $\eta_{\parallel}$\,, $\eta_{\perp}$ originates from the expansion of a suitably defined vector $\eta^I$ in the $(T,N)$ basis \cite{CAP}. Note that $\eta_{\perp}$ is a dimensionless turning rate. In the context of Gauss-Bonnet inflation, one can introduce the following additional slow-roll parameters:
\be \label{SR_GB}
\delta_1 = H \dot{f} \qquad , \qquad \delta_2 = \frac{\ddot{f}}{H \dot{f}}  \qquad ;
\ee
see, for example \cite{MFMM} and references therein. 

In terms of the above parameters, (\ref{Hd}) becomes:
\be
2 \dot{H} = - \dot{\sigma}^2 + \delta_1 H^2 + 2 \delta_1 \varepsilon H^2 - \delta_1 \delta_2 H^2 \quad ,
\ee
or equivalently:
\be \label{sigma-d}
\frac{\dot{\sigma}^2}{2 H^2} = \varepsilon + \frac{\delta_1}{2} + \delta_1 \varepsilon - \frac{\delta_1 \delta_2}{2} \,\,\, . 
\ee
Also, (\ref{gEoM}) acquires the form:
\be \label{H2V}
3 H^2 = \frac{\dot{\sigma}^2}{2} + V - \,3 \delta_1 H^2 \,\,\, .
\ee

\subsection{Slow-roll approximation}

We will consider a slow-roll regime defined by the following conditions:
\be \label{SRdef}
\varepsilon \ll 1 \quad , \quad |\eta_{\parallel}| \ll 1 \quad , \quad |\delta_1| \ll 1 \quad , \quad |\delta_2| \ll 1 \quad .
\ee
The work of Koh \cite{SK} imposes additional slow-roll conditions. However, we will see below that these conditions are too restrictive in the rapid-turn regime of interest here.

In leading order, within the approximations (\ref{SRdef}), the relation (\ref{sigma-d}) becomes:
\be \label{sigmaAppr}
\frac{\dot{\sigma}^2}{H^2} \approx 2 \varepsilon + \delta_1 \,\,\, .
\ee
Thus, during slow-roll (\ref{H2V}) gives:
\be \label{HsqV}
3 H^2 \approx V \,\,\, ,
\ee
just like in the absence of the Gauss-Bonnet term.

In the slow-roll approximation defined above, (\ref{Omf}) can be written as:
\be \label{OmAppr}
\Omega \approx \frac{\left( V_N - 3 f_N H^4 \right)}{\dot{\sigma}} \approx \frac{\left(V_N - \frac{1}{3} f_N V^2 \right)}{\dot{\sigma}} \,\,\, .
\ee
This suggests that, through a suitable $f_N$\,, the Gauss-Bonnet term can contribute to (or even drive) field-space trajectories with a (suitably) large turning rate. In that regard, note that neither of the two terms in (\ref{OmAppr}) has a definite sign. So the $V_N$ and $f_N$ contributions may cancel each other or may enhance each other, depending on specific models and background solutions.

\subsection{Rapid-turn regime}

The rapid-turn approximation was originally defined by the condition \cite{CAP,AP,BFM}:
\be \label{RTc}
\eta_{\perp}^2 \gg 1 \,\,\, .
\ee
Recently, a more general (weaker) version of the rapid turn regime, defined by \cite{TB}:
\be \label{RTcw}
\eta_{\perp}^2 \, \gg \, \varepsilon  \,, \,|\eta_{\parallel}| \,, \,|\delta_1| \,, \,|\delta_2| \,\,\, ,
\ee
has also been studied in the literature. Unless otherwise specified, our considerations in subsequent Sections will require only (\ref{RTcw}).

Note that during slow-roll, in view of (\ref{sigmaAppr})-(\ref{OmAppr}), the condition (\ref{RTc}) has the form:
\be
\frac{(V_N-\frac{1}{3} f_N V^2)^2}{V^2} \gg \frac{1}{9} (2 \varepsilon + \delta_1)
\ee
or equivalently
\be \label{RTc_f}
\frac{(V_N-3 f_N H^4)^2}{H^4} \gg (2 \varepsilon + \delta_1) \,\,\, .
\ee
Similarly, (\ref{RTcw}) implies that:
\be \label{RTcw_f}
\frac{(V_N-3 f_N H^4)^2}{H^4} \gg (2 \varepsilon + \delta_1) \varepsilon \,\,\, ,
\ee
for the condition $\eta_{\perp}^2 \gg \varepsilon$ etc. This inequality can be satisfied even for $V_N = 0$\,, in which case it reduces to $9 f_N^2 H^4 \gg (2 \varepsilon + \delta_1) \varepsilon$\,. Restoring explicitly $M_P$, the last relation can be written as $9 (f_N H)^2 (H/M_P)^2 \gg (2 \varepsilon + \delta_1) \varepsilon $\,. Note that the combination $f_N H$ is dimensionless, while the ratio $H/M_P$ can be bounded by the current lack of observational evidence for primordial gravitational waves.

To obtain novel effects due to the Gauss-Bonnet term, in the following we will treat $f_N H^2$ on par with $V_N$ in the rapid term regime. This is in contrast with the approximation considered in \cite{SK}, where $f_N H^2$ was treated on par with the slow-roll parameters.

\section{Relation between $\varepsilon$ and $\varepsilon_V$} \label{EpEpV}
\setcounter{equation}{0}

In this section we will derive the relation between the $\varepsilon$-parameter defined in terms of $H(t)$ in (\ref{Param_def}) and its counterpart defined in terms of the potential $V$ as:
\be \label{ep_V}
\varepsilon_V = \frac{V_I V^I}{2 V^2} \,\,\, .
\ee
In single-field inflation, to leading order during slow-roll, one has $\varepsilon \approx \varepsilon_V$\,. It is well-known that in the standard multifield context this relation is modified, even within the slow-roll approximation. Namely, for two scalars, one has:
\be \label{EpV-Ep}
\varepsilon_V \, \approx \, \varepsilon \left( 1 + \frac{\eta_{\perp}^2}{9} \right) \,\,\, ,
\ee
where $\eta_{\perp}$ is the dimensionless turning rate introduced in (\ref{Param_def}). This was discussed in detail in \cite{AP}, in the context of evading the swampland conjectures. Following that discussion, we derive the further modification that the Gauss-Bonnet term induces.

Let us begin by rewriting (\ref{ep_V}) in the basis (\ref{basis}):
\be \label{Ep_Vn_Vt}
\varepsilon_V = \frac{V_T^2}{2V^2} + \frac{V_N^2}{2V^2} \,\,\, .
\ee
The second term is easier to compute. So let us consider it first. From (\ref{Omf}), we have:
\be \label{VN_om}
V_N = \dot{\sigma} \Omega + 3 f_N H^4 (1 - \varepsilon) \,\,\, ,
\ee
whereas (\ref{sigma-d})-(\ref{H2V}) imply that:
\be \label{V_SR}
V = H^2 \left( 3 - \varepsilon + \frac{5}{2} \delta_1 - \delta_1 \varepsilon + \frac{1}{2} \delta_1 \delta_2 \right) \,\,\, .
\ee
Combining (\ref{VN_om}) and (\ref{V_SR}) gives:
\be
\frac{V_N^2}{2V^2} \,= \,\frac{\left[ \dot{\sigma} \Omega + 3 f_N H^4 (1 - \varepsilon )\right]^2}{2 H^4 \!\left( 3 - \varepsilon + \frac{5}{2} \delta_1 - \delta_1 \varepsilon + \frac{1}{2} \delta_1 \delta_2 \right)^2} \,\,\, .
\ee
Hence, keeping only the leading terms in the slow-roll approximation, we find:
\be \label{VN}
\frac{V_N^2}{2V^2} \,\approx \,\frac{1}{2} \left( \frac{\sqrt{(2 \varepsilon + \delta_1)} \,\eta_{\perp}}{3} + f_N H^2 \right)^{\!2} \,\,\, ,
\ee
where we have used (\ref{sigmaAppr}).

Now we turn to computing the first term in (\ref{Ep_Vn_Vt}). For that purpose, let us differentiate the slow-roll relation (\ref{sigmaAppr}) with respect to time. This gives:
\be \label{SR-dif}
\frac{\dot{\sigma} \ddot{\sigma}}{H^2} - \frac{\dot{\sigma}^2 \dot{H}}{H^3} \approx \dot{\varepsilon} + \frac{\dot{\delta}_1}{2} \,\,\, .
\ee
From (\ref{sigma-dd}) we have:
\be \label{sigma-dd-s}
\ddot{\sigma} = - 3 H \dot{\sigma} - V_T + 3 f_T H^4 (1 - \varepsilon ) \,\,\, ,
\ee
whereas differentiating the definition $\delta_1 = H \dot{f}$ gives:
\be \label{delta1-d}
\dot{\delta}_1 = \dot{H} \dot{f} + H \ddot{f} = H ( - \varepsilon \delta_1 + \delta_1 \delta_2 ) \,\,\, . 
\ee
Substituting (\ref{sigma-dd-s}) and (\ref{delta1-d}) in (\ref{SR-dif}), as well as using (\ref{sigmaAppr}) and the definition
\be
\eta = \frac{\dot{\varepsilon}}{H \varepsilon} \,\,\, ,
\ee
we find:
\be
\frac{\dot{\sigma} \left[ - 3 H \dot{\sigma} - V_T + 3 f_T H^4 (1 - \varepsilon )\right]}{H^2} - (2 \varepsilon + \delta_1) \frac{\dot{H}}{H} \, \approx \, H \varepsilon \eta + \frac{H}{2} ( - \varepsilon \delta_1 + \delta_1 \delta_2 ) \,\,\, . 
\ee
Solving this for $V_T$ gives:
\be \label{VT_prelim}
V_T \,\approx \,- \frac{H^2}{\dot{\sigma}} \left( \frac{3 \dot{\sigma}^2}{H} - 3 \dot{\sigma} f_T H^2 (1- \varepsilon) + \frac{2 \varepsilon \dot{H}}{H} + \frac{\delta_1 \dot{H}}{H} + H \varepsilon \eta - \frac{H}{2} \varepsilon \delta_1 + \frac{H}{2} \delta_1 \delta_2 \right) \,.
\ee
Notice that $H \dot{\sigma} f_T = H \dot{\sigma} f_I T^I = H f_I \dot{\phi}_0^I = H \dot{f} = \delta_1$\,. Thus (\ref{VT_prelim}) implies:
\be \label{VT_prelim2}
V_T^2 \,\approx \,\frac{H^4}{2 \varepsilon + \delta_1} \left( 6 \varepsilon + \frac{3}{2} \varepsilon \delta_1 - 2 \varepsilon^2 + \varepsilon \eta + \frac{\delta_1 \delta_2}{2} \right)^{\!2} \, ,
\ee
where we have used (\ref{sigmaAppr}) again. Combining (\ref{VT_prelim2}) with (\ref{V_SR}) and keeping only the leading terms in the slow-roll approximation, we finally obtain:
\be \label{VT}
\frac{V_T^2}{2 V^2} \, \approx \, \frac{2\varepsilon^2}{(2\varepsilon + \delta_1)} \,\,\, .
\ee

In conclusion, substituting (\ref{VT}) and (\ref{VN}) in (\ref{Ep_Vn_Vt}), we find:
\be \label{Ep_V_1}
\varepsilon_V \,\, \approx \,\, \frac{2\varepsilon^2}{(2\varepsilon + \delta_1)} \,+ \,\frac{1}{2} \!\left( \frac{\sqrt{\left( 2 \varepsilon + \delta_1 \right)} \,\eta_{\perp}}{3} + f_N H^2 \right)^{\!2} \, ,
\ee
to leading order during slow-roll. Note that for $f (\phi) = const$, which implies that both $\delta_1 = 0$ and $f_N = 0$\,, we recover the familiar result (\ref{EpV-Ep}). However, (\ref{Ep_V_1}) shows that even for non-large turning rate $|\eta_{\perp}|$\,, one can have $\varepsilon_V \sim {\cal O} (1)$ or larger, together with small $\varepsilon$\,, depending on $\delta_1$ and $f_N = N_I f^I$. This opens up new possibilities for satisfying the swampland conjectures, while having inflation with $\varepsilon \ll 1$\,.

\section{Relation between the Hessians of $V$ and $f$} \label{HesVf}
\setcounter{equation}{0}

In single-field inflationary models there is another important slow-roll parameter, defined via the potential, namely: $\eta_V = \frac{V''}{V}$\,. Its generalization to the multifield context is more subtle than in the case of $\varepsilon_V$\,. One option, discussed in the literature, is to consider the minimal eigenvalue of the (suitably normalized) Hessian of the potential: $\eta_{IJ} \equiv \frac{\nabla_I \nabla_J V}{V}$ \cite{ACPRZ}. Another is to define $\eta_V$ as $\frac{{{\rm Tr} (\eta_{IJ})}}{N_f}$\,, where $N_f$ is the number of scalar fields \cite{CPXZ}. In any case, the relationship between $\eta_{\parallel}$ (the standard $\eta$-parameter in single-scalar models) and the second derivatives of the potential is complicated and allows for $|\eta_{\parallel}| \ll 1$\,, together with non-small components of $\eta_{IJ}$\,. In this section we derive the modifications to this relation, due to the inclusion of the Gauss-Bonnet term. 

\subsection{Magnitude of $\nabla_T \nabla_T f$} \label{fTT_mag}

To properly account for the contribution of $f$-dependent terms to the Hessian of $V$\,, let us first consider the implications of the slow-roll and rapid turn approximation for the derivatives of the coupling function. In view of the discussion in Section \ref{SR-RT}, we will consider terms containing $f_N$ as non-negligible, while terms with $f_T$ as slow-roll suppressed. At first sight, this might suggest that $\nabla_T \nabla_T f \equiv T^I T^J \nabla_I \nabla_J f$ is a small quantity during slow-roll (an assumption that was made in \cite{SK}.) However, here we will show that it is non-negligible in the rapid turn and slow-roll regime and that it can, even, be {\it large} during {\it slow turn} and slow-roll. For convenience, we will use the notation:\footnote{In \cite{SK} $\nabla_T \nabla_T f$ is denoted by $f_{TT}$ and similarly for $f_{TN}$ and $f_{NN}$\,. However, here we always denote by a lower index on a function just an ordinary derivative of the function.} 
\be
f_{;\,TT} \equiv \nabla_T \nabla_T f = T^I T^J \nabla_I \nabla_J f \,\,\, .
\ee

Now, let us compute:
\be \label{fdd}
\ddot{f} = \frac{d}{dt} (\dot{f}) = \frac{d}{dt} (f_I \dot{\phi}_0^I) = f_{IJ} \dot{\phi}_0^I \dot{\phi}_0^J + f_I \ddot{\phi}_0^I \,\,\, .
\ee
From (\ref{sEoM})-(\ref{Defs}), we have:
\be
\ddot{\phi}_0^I = - \Gamma^I_{JK} \dot{\phi}_0^J \dot{\phi}_0^K - 3 H \dot{\phi}_0^I - V^I + 3 f^I H^4 (1 - \varepsilon) \,\,\, .
\ee
Substituting this into (\ref{fdd}), as well as using that $\nabla_I \nabla_J f = f_{IJ} - \Gamma^K_{IJ} f_K$\,, gives:
\be
\ddot{f} = (\nabla_I \nabla_J f) \,\dot{\phi}_0^I \dot{\phi}_0^J - 3 H \dot{f} - f_I V^I + 3 f_I f^I H^4 (1 - \varepsilon) \,\,\, .
\ee
This can be rewritten as:
\be
\delta_1 \delta_2 = \dot{\sigma}^2 \nabla_T \nabla_T f - 3 \delta_1 - f_T V_T - f_N V_N + 3 (f_T^2 + f_N^2) H^4 ( 1 - \varepsilon) \,\,\, .
\ee
Solving the last relation for $f_{;\,TT} = \nabla_T \nabla_T f$\,, we find:
\be \label{fTT-prelim}
f_{;\,TT} = \frac{1}{\dot{\sigma}^2} \left[ \delta_1 \delta_2 + 3 \delta_1 + f_T V_T + f_N V_N - 3 (f_T^2 + f_N^2) H^4 (1 - \varepsilon) \right] \,\,\, .
\ee
Now, (\ref{sigma-dd}) implies:
\be
V_T = H \dot{\sigma} \left( \eta_{\parallel} - 3 + 3 \frac{f_T}{\dot{\sigma}} H^3 (1 - \varepsilon)\right) \,\,\, ,
\ee
where we have used (\ref{Param_def}). Substituting this in (\ref{fTT-prelim}), we obtain:
\bea \label{fTT-prelim2}
f_{;\,TT} &=& \frac{1}{\dot{\sigma}^2} \left[ \delta_1 \delta_2 + 3 \delta_1 + \delta_1 \eta_{\parallel} - 3 H \dot{\sigma} + f_N V_N - 3 f_N^2 H^4 (1 -\varepsilon) \right] \nn \\ 
&=& \frac{1}{\dot{\sigma}^2} \left[ \delta_1 \delta_2 + 3 \delta_1 + \delta_1 \eta_{\parallel} - 3 H \dot{\sigma} + f_N \Omega \dot{\sigma} \right] \,\, ,
\eea
where we have used (\ref{Omf}). Neglecting the second order (in slow-roll) terms in the numerator of (\ref{fTT-prelim2}) and using (\ref{sigmaAppr}) gives:
\be \label{fTT-1st}
f_{;\,TT} \approx \frac{3 \delta_1}{H^2 (2 \varepsilon + \delta_1)} + \frac{(f_N \eta_{\perp} - 3)}{\sqrt{2 \varepsilon + \delta_1}} \,\,\, .
\ee 
Hence the leading term in the slow-roll and rapid turn approximation is:
\be
f_{;\,TT} \approx \frac{f_N \,\Omega}{\dot{\sigma}} = \frac{f_N \eta_{\perp}}{\sqrt{2 \varepsilon + \delta_1}} \,\,\, .
\ee
This is not slow-roll suppressed, since for us the quantity \,$f_N \eta_{\perp}$ is not small. Furthermore, $|f_{;\,TT}|$ would not be small even if the inequality $|f_N \eta_{\perp}| \ll 1$ were satisfied. Indeed, in such a case, according to (\ref{fTT-1st}), the leading contribution would be $f_{;TT} \approx - \frac{3}{\sqrt{2 \varepsilon + \delta_1}}$\,, which can have a large magnitude for sufficiently small $\varepsilon$ and $|\delta_1|$\,. This means that the approximation $H^2 f_{;TT} \ll 1$\,, assumed in \cite{SK} as one of the slow-roll conditions, is not only invalid in a rapid turn regime, but may even be violated in the {\it slow-turn} slow-roll approximation. 

In view of the above result, in the following we will treat $\nabla_N \nabla_T f$ and $\nabla_N \nabla_N f$ as non-negligible too, even though for them no expressions in the vein of (\ref{fTT-1st}) can be derived.

\subsection{Relation between $\nabla_T \nabla_T V$ and $\nabla_T \nabla_T f$}

Now we will consider the components of the Hessian of the potential one by one. We begin with $\nabla_T \nabla_T V \equiv T^I T^J \nabla_I \nabla_J V$\,. For convenience, let us denote:
\be
V_{;\,TT} \equiv \nabla_T \nabla_T V = T^I T^J \nabla_I \nabla_J V \qquad .
\ee
We can write this as:
\bea \label{VTT}
V_{;\,TT} &=& T^I \nabla_I (T^J V_J) - T^I (\nabla_I T^J) V_J = \nabla_T (V_T) - \frac{1}{\dot{\sigma}} (D_t T^J) V_J \nn \\
&=& \nabla_T (V_T) + \frac{1}{\dot{\sigma}} \Omega N^J V_J = \nabla_T (V_T) + H^2 \eta_{\perp}^2 + \frac{1}{\dot{\sigma}} 3 \eta_{\perp} f_N H^5 (1 - \varepsilon) \,\,\, , 
\eea
where we have used (\ref{TN-Om}), (\ref{Omf}) and (\ref{Param_def}). Similarly, we compute:
\be \label{fTT}
f_{;\,TT} = \nabla_T (f_T) - \frac{1}{\dot{\sigma}} (D_t T^J) f_J = \nabla_T (f_T) + \frac{1}{\dot{\sigma}} \Omega N^J f_J = \nabla_T (f_T) + \frac{1}{\dot{\sigma}} H \eta_{\perp} f_N \,\,\, .
\ee

One can relate $V_{;\,TT}$ and $f_{;\,TT}$ via the scalar field equations. Namely, let us act with $\nabla_T = \frac{\dot{\phi}_0^I}{\dot{\sigma}} \nabla_I = \frac{1}{\dot{\sigma}} \frac{d}{dt}$ on (\ref{sigma-dd}). This gives:
\be
\hspace*{-0.2cm}\frac{\dddot{\sigma}}{\dot{\sigma}} + 3 \dot{H} + 3 H \frac{\ddot{\sigma}}{\dot{\sigma}} + \nabla_T (V_T) - 3 [\nabla_T (f_T)] H^4 (1-\varepsilon) - \frac{12 f_T H^3 \dot{H} (1-\varepsilon)}{\dot{\sigma}} + \frac{3 f_T H^4 \dot{\varepsilon}}{\dot{\sigma}} = 0 \,\,\, .
\ee
Using the slow-roll parameters $\xi \equiv \frac{\dddot{\sigma}}{H^2 \dot{\sigma}}$ and $\eta = \frac{\dot{\varepsilon}}{H \varepsilon}$\,, as well as that $f_T = \frac{\dot{f}}{\dot{\sigma}}$\,, the last relation can be rewritten as:
\be
\nabla_T (V_T) - 3 H^4 (1-\varepsilon) \nabla_T (f_T) = 3 H^2 \!\left( \varepsilon + \eta_{\parallel} - \frac{\xi}{3} \right) - \frac{12 H^4 \delta_1 \varepsilon (1 - \varepsilon)}{\dot{\sigma}^2} - \frac{3 H^3 \delta_1 \eta \varepsilon}{\dot{\sigma}^2} \,\,.
\ee
In the slow-roll approximation this becomes:
\be \label{VTfT}
\nabla_T (V_T) - 3 H^4 (1-\varepsilon) \nabla_T (f_T) \approx 3 H^2 \!\left( \varepsilon + \eta_{\parallel} - \frac{\xi}{3} \right) - \frac{12 H^2 \delta_1 \varepsilon}{(2 \varepsilon + \delta_1)} \,\,\, ,
\ee
where we have used (\ref{sigmaAppr}) and dropped second order terms in the slow-roll parameters. 

Now, substituting in (\ref{VTfT}) the expressions for $\nabla_T (V_T)$ from (\ref{VTT}) and for $\nabla_T (f_T)$ from (\ref{fTT}), we finally find to leading order:
\be
V_{;\,TT} \approx 3 H^4 f_{;\,TT} + H^2 \eta_{\perp}^2 \,\,\, .
\ee
Note that for $f (\phi) = const$ this reduces to the standard multifield result \cite{AGHPP}:
\be
V_{;\,TT} \approx H^2 \eta_{\perp}^2 \,\,\, ,
\ee 
at leading order during rapid turn and slow-roll. However, in view of (\ref{fTT-1st}), when $f_{;\,TT} \neq 0$ the expression for $V_{;\,TT}$ gets modified even in a {\it slow turn} and slow-roll regime.

\subsection{Relation between $\nabla_T \nabla_N V$ and $\nabla_T \nabla_N f$}

Let us now derive the relationship between $\nabla_T \nabla_N V \equiv T^I N^J \nabla_I \nabla_J V$ and $\nabla_T \nabla_N f$\,, defined in the same way. For convenience, we introduce the notation:
\be
V_{;\,TN} \equiv \nabla_T \nabla_N V \qquad {\rm and} \qquad f_{;\,TN} \equiv \nabla_T \nabla_N f \quad .
\ee

In the slow-roll approximation, we have from (\ref{Omf}):
\be
V_N - 3 f_N H^4 = \dot{\sigma} \Omega \,\,\, .
\ee
Acting with $\nabla_T \equiv T^I \nabla_I$ on both sides of this relation, we find:
\be \label{NT}
\nabla_T (V_N - 3 f_N H^4) = \frac{1}{\dot{\sigma}} \frac{d}{dt} (\dot{\sigma} \Omega) \,\,\, .
\ee
The right hand side gives:
\be \label{NT_rhs}
\frac{1}{\dot{\sigma}} \frac{d}{dt} (\dot{\sigma} \Omega) = - H^2 \eta_{\perp} (\eta_{\parallel} + \varepsilon + \nu) \,\,\, ,
\ee
where $\nu \equiv - \frac{\dot{\eta}_{\perp}}{H \eta_{\perp}}$ is another small parameter.\footnote{The requirement $|\nu| \ll 1$ ensures that the rapid turn regime can be sustained for a prolonged period of time.}

The left hand side of (\ref{NT}) can be written as:
\bea \label{nTVN}
\nabla_T (V_N - 3 f_N H^4) &=& T^I \nabla_I V_N - 3 T^I \nabla_I (f_N H^4) \nn \\
&=& T^I \nabla_I (N^J V_J) - 3 H^4 \nabla_T (f_J N^J) - 3 f_N T^I \nabla_I H^4 \nn \\
&=& V_{;\,TN} + T^I (\nabla_I N^J) V_J - 3 H^4 f_{;\,TN} - 3 H^4 T^I (\nabla_I N^J) f_J - \frac{3 f_N }{\dot{\sigma}} \frac{d}{dt} H^4 \nn \\
&=& V_{;\,TN} + \frac{1}{\dot{\sigma}} (D_t N^J) V_J - 3 H^4 f_{;\,TN} - \frac{3 H^4}{\dot{\sigma}} (D_t N^J) f_J - \frac{12 f_N H^3 \dot{H}}{\dot{\sigma}} \nn \\
&\approx& V_{;\,TN} + \frac{\Omega V_T}{\dot{\sigma}} - 3 H^4 f_{;\,TN} - \frac{3 H^4 \Omega f_T}{\dot{\sigma}} + \frac{12 H^4 f_N \varepsilon}{\sqrt{2 \varepsilon + \delta_1}} \quad ,
\eea
where we have used (\ref{TN-Om}) and (\ref{sigmaAppr}).
Note that, due to (\ref{sigma-dd}) and (\ref{Param_def}), one has:
\be \label{OmVT}
\frac{\Omega V_T}{\dot{\sigma}} = \frac{\Omega}{\dot{\sigma}} H \dot{\sigma} \left( \eta_{\parallel} - 3 + 3 \frac{f_T H^3}{\dot{\sigma}} \right) = \Omega H (\eta_{\parallel} - 3) + \frac{3H^4\Omega f_T}{\dot{\sigma}} \,\,\, ,
\ee
where we have neglected the $\varepsilon$-correction in the $f_T$-term. Substituting (\ref{OmVT}) in (\ref{nTVN}), we finally find:
\be \label{NTVNfN}
\nabla_T (V_N - 3 f_N H^4) = V_{;\,TN} + H^2 \eta_{\perp} (\eta_{\parallel} - 3) - 3 H^4 f_{;\,TN} + \frac{12 H^4 f_N \varepsilon}{\sqrt{2 \varepsilon + \delta_1}} \,\,\, .
\ee

Combining (\ref{NT})-(\ref{NT_rhs}) and (\ref{NTVNfN}) implies that:
\be \label{VTN}
V_{;\,TN} \approx 3 H^4 f_{;\,TN} + 3 H^2 \eta_{\perp} \,\,\, ,
\ee
to leading order during slow-roll and rapid turn. In writing (\ref{VTN}), we have assumed that the slow-roll parameters are sufficiently small, such that $\frac{\varepsilon}{\sqrt{2 \varepsilon + \delta_1}} \ll 1$\,. Otherwise, the last term in (\ref{NTVNfN}) may not be negligible at leading order. Note that for $f (\phi^I) = const$ we recover the familiar leading-order result \cite{AGHPP}:
\be
V_{;\,TN} \approx 3 H^2 \eta_{\perp} \,\,\, .
\ee

\subsection{Summary and discussion}

We have found, to leading order in the slow-roll and rapid turn approximation, the following relations for the Hessian components of the potential:
\be \label{VfTT_VfTN}
V_{;\,TT} \approx 3 H^4 f_{;\,TT} + H^2 \eta_{\perp}^2 \qquad , \qquad V_{;\,TN} \approx 3 H^4 f_{;\,TN} + 3 H^2 \eta_{\perp} \quad .
\ee
Note that all of the quantities $f_{;\,TT}$\,, $f_{;\,TN}$ and $\eta_{\perp}$ can have either sign. So the $f$-terms can enhance the $\eta_{\perp}$-terms or can subtract from them, depending on specific models. Interestingly, this opens the possibility to realize a rapid turn regime even with $V_{;\,TT} \approx 0$ and/or $V_{;\,TN} \approx 0$\,, in contrast with the standard multifield context without the Gauss-Bonnet term. It is also notable that, since $f_{;TT}$ can be large even during a {\it slow turn} regime as we saw in Subsection \ref{fTT_mag}, so can $V_{;TT}$ due to (\ref{VfTT_VfTN}).

In view of the above, the normalized Hessian components are:
\be \label{VfTT_VfTN_V}
\frac{V_{;\,TT}}{V} \, \approx \, H^2 \,f_{;\,TT} \,+ \,\frac{\eta_{\perp}^2}{3} \qquad , \qquad \frac{V_{;\,TN}}{V} \, \approx \, H^2 \,f_{;\,TN} \,+ \,\eta_{\perp} \quad ,
\ee
to leading order. However, it is not possible to express $\frac{V_{;\,NN}}{V}$ exclusively in terms of kinematic quantities and derivatives of $f(\phi)$. Nevertheless, we can draw some interesting conclusions about the difference with the single-scalar case. Indeed, in single-field slow-roll inflation there is a simple relation between the second slow-roll parameter, $\eta_{\parallel}$ according to our notation in (\ref{Param_def}), and its potential counterpart $\eta_V \equiv V''/V$:
\be
\eta_V \approx \eta_{\parallel} + \varepsilon \,\,\, . 
\ee
This implies that small $|\eta_{\parallel}|$ and $\varepsilon$ require small $|\eta_V|$\,, thus leading to the famous $\eta$-problem. On the other hand, in multifield inflation $\eta_V$ can be defined as the minimum eigenvalue of the matrix $V_{;IJ}/V$. For two fields, it depends on $V_{;TT}$, $V_{;TN}$, and $V_{;NN}$. The first two, as noted above, can be expressed in terms of kinematic quantities, while the latter cannot. Despite that, information about $\eta_V$ can be obtained from the Rayleigh quotient, which implies that $\eta_V < \frac{V_{;TT}}{V}$ and $\eta_V < \frac{V_{;NN}}{V}$. These inequalities were first identified in \cite{Chakraborty:2019dfh}, where it was emphasized that rapid-turn inflation does not suffer from an $\eta$-problem, since in this regime $\eta_V \sim {\cal O} (1)$ is consistent with slow-roll. The Gauss-Bonnet term’s contribution to $V_{;TT}$ may be positive or negative, thereby affecting the degree of turning required to resolve the $\eta$-problem.

\section{Consistency condition for SR and RT} \label{ConsC-SRRT}
\setcounter{equation}{0}

In standard multifield cosmology, without a Gauss-Bonnet term, one can derive a consistency condition, for two fields, relating the scalar potential and field-space metric, which is necessary for the existence of slow-roll and rapid-turn inflationary trajectories \cite{TB, AL, LA, WIA} \footnote{The consistency condition for more than two fields was derived in \cite{Aragam:2020uqi}}. Here we will derive the modification of this condition due to the Gauss-Bonnet term.

An essential ingredient in the considerations of \cite{TB, AL, LA, WIA} is the potential gradient basis in field space:
\be \label{nt_def}
n^I = \frac{V^I}{\sqrt{G^{JK} V_J V_K}} \qquad , \qquad
\tau_I = \sqrt{\det G} \,\epsilon_{IJ} \,n^J \quad ,
\ee
which was introduced first in \cite{BM}. This basis does not depend on any inflationary trajectory, which makes it very useful for addressing general questions without knowledge of specific solutions. It can be related to the kinematic basis (\ref{basis}) via a rotation by an angle that depends both on time and on the field space trajectory. Using the basis (\ref{nt_def}) and the background equations of motion, one can derive simple expressions for $\eta_{\perp}$ and $\eta_{\parallel}$\,, which play a key role in the derivation of the slow-roll and rapid-turn consistency condition in \cite{AL,LA}. To follow the same steps as in \cite{LA}, and in view of the background equations (\ref{sEoM}), we now introduce the following new basis in field space:
\be \label{nt_hat_def}
{\hat n}^I = \frac{V^I - 3 f^I H^4 (1 - \varepsilon)}{\sqrt{G^{JK} \left[ V_J - 3 f_J H^4 (1-\varepsilon) \right] \!\left[V_K - 3 f_K H^4 (1-\varepsilon)\right]}} \quad , \quad
{\hat \tau}_I = \sqrt{\det G} \,\epsilon_{IJ} \,{\hat n}^J \,\, .
\ee
In the slow-roll approximation, using (\ref{SRdef}) and (\ref{HsqV}), we have:
\be \label{n_hat_SR}
{\hat n}^I \approx \frac{V^I - \frac{1}{3} f^I V^2}{\sqrt{G^{JK} \!\left( V_J - \frac{1}{3} f_J V^2 \right) \!\left( V_K - \frac{1}{3} f_K V^2 \right)}} \quad .
\ee
Thus, in the regime of interest for us, the basis $(\hat{n}, \hat{\tau})$ is determined by the field-space metric, scalar potential and coupling function (i.e., it is trajectory-independent), in analogy with the potential gradient basis $(n , \tau)$\,. And, of course, for $f (\phi^I) = const$, the new basis reduces exactly to (\ref{nt_def}).

The kinematic basis (\ref{basis}) can be related to the new basis (\ref{nt_hat_def}) via a rotation:
\bea
\label{BasesRel}
T \!&=& \!\cos \vartheta \,\, \hat{n} \,+ \,\sin \vartheta \,\, \hat{\tau} \,\,\, , \nn \\
N \!&=& \!-\sin \vartheta \,\, \hat{n} \,+ \,\cos \vartheta \,\, \hat{\tau} \,\,\, ,
\eea
where the angle $\vartheta$ depends on time, as well as on the field space trajectory. Using (\ref{sigma-dd}) and (\ref{Omf}), together with (\ref{BasesRel}) and (\ref{nt_hat_def}), one finds:
\be \label{eta_par_cos}
\eta_{\parallel} \,= \,- \frac{\ddot{\sigma}}{H \dot{\sigma}} \,= \,3 + \frac{T^I [ V_I - 3 f_I H^4 (1-\varepsilon) ]}{H \dot{\sigma}} \,= \,3 + \frac{\cos \vartheta}{c} \quad ,
\ee
and
\be \label{eta_perp_sin}
\eta_{\perp} \,= \,\frac{N^I [V_I - 3 f_I H^4 (1-\varepsilon)]}{H \dot{\sigma}} \,= \,- \frac{\sin \vartheta}{c} \quad ,
\ee
where
\be \label{c_def}
c \,\equiv \frac{H \dot{\sigma}}{\sqrt{G^{JK} [V_J - 3 f_J H^4 (1-\varepsilon)] [V_K - 3 f_K H^4(1-\varepsilon)]}} \quad .
\ee
For $f (\phi) = const$\,, relations (\ref{eta_par_cos})-(\ref{c_def}) reduce precisely to the corresponding expressions in \cite{AL}, obtained in the absence of the Gauss-Bonnet term. From (\ref{eta_par_cos})-(\ref{eta_perp_sin}) it follows that:
\be
\eta_{\perp}^2 + (\eta_{\parallel} - 3)^2 = \frac{1}{c^2} \,\,\, ,
\ee
which implies, in particular, that during slow-roll:
\be \label{eta_c}
\eta_{\perp}^2 \approx \frac{1}{c^2} - 9 \,\,\, .
\ee
Thus, the regime $\eta_{\perp}^2 \gg 1$ corresponds to $c^2 \ll 1$\,. In the following, we will not assume that $c^2$ is small, though. Hence our results will apply in the general case (\ref{RTcw}).

We now derive a consistency condition for slow-roll and rapid-turn inflation, following the steps outlined in \cite{LA}. For that purpose, let us introduce the notation:
\be \label{ModHess}
{\cal V}_{\hat{n} \hat{n}} \equiv V_{\hat{n} \hat{n}} - \frac{1}{3} f_{\hat{n} \hat{n}} V^2 \quad , \quad {\cal V}_{\hat{n} \hat{\tau}} \equiv V_{\hat{n} \hat{\tau}} - \frac{1}{3} f_{\hat{n} \hat{\tau}} V^2 \quad , \quad {\cal V}_{\hat{\tau} \hat{\tau}} \equiv V_{\hat{\tau} \hat{\tau}} - \frac{1}{3} f_{\hat{\tau} \hat{\tau}} V^2 \quad .
\ee
where \,$V_{\hat{n} \hat{n}} = \hat{n}^I \hat{n}^J \nabla_I V_J$\,\,, \,$f_{\hat{n} \hat{n}} = \hat{n}^I \hat{n}^J \nabla_I f_J$ \,etc. Since we are working in the slow-roll approximation, here and in the following we can view $\hat{n}$ as given by (\ref{n_hat_SR}). Using (\ref{BasesRel}), we compute:
\bea
\label{VTN-Vntau}
\hspace*{-0.4cm}V_{;\,TT} - \frac{1}{3} f_{;\,TT} V^2 &=& {\cal V}_{\hat{n} \hat{n}} \,\cos^2 \vartheta \,+ \,2 \,{\cal V}_{\hat{n} \hat{\tau}} \,\sin \vartheta \,\cos \vartheta \,+ \,{\cal V}_{\hat{\tau} \hat{\tau}} \,\sin^2 \vartheta \,\,\, , \nn \\
\hspace*{-0.4cm}V_{;\,TN} - \frac{1}{3} f_{;\,TN} V^2 &=& \left( {\cal V}_{\hat{\tau} \hat{\tau}} - {\cal V}_{\hat{n} \hat{n}} \right) \,\cos \vartheta \,\sin \vartheta \,+ \,{\cal V}_{\hat{n} \hat{\tau}} \left( \cos^2 \vartheta - \sin^2 \vartheta \right) \,\, .
\eea
In the slow-roll approximation, this becomes:
\bea
\label{VTN-Vntau_V_c}
\frac{V}{3\,c^2} - 3 V &\approx & 9 c^2 {\cal V}_{\hat{n} \hat{n}} - 6 s c \,\sqrt{1-9c^2} \,{\cal V}_{\hat{n} \hat{\tau}} + (1-9c^2) {\cal V}_{\hat{\tau} \hat{\tau}} \,\,\, , \nn \\
\frac{s}{c} \sqrt{1-9c^2}\,V &\approx & 3 s c \,\sqrt{1-9c^2} \left( {\cal V}_{\hat{\tau} \hat{\tau}} - {\cal V}_{\hat{n} \hat{n}} \right) + (1-18c^2) {\cal V}_{\hat{n} \hat{\tau}} \,\,\, ,
\eea
where we have used (\ref{VfTT_VfTN_V}), (\ref{HsqV}) and (\ref{eta_c}), as well as that due to (\ref{eta_par_cos}) during slow-roll one has: $\cos \vartheta \approx - 3c$ and $\sin \vartheta \approx s \sqrt{1 - 9 c^2}$ with $s = \pm 1$\,. This system has exactly the same form as the system (3.4) in \cite{LA}, obtained in the absence of the Gauss-Bonnet term. Thus, performing on (\ref{VTN-Vntau_V_c}) the same manipulations, as described in \cite{LA}, gives:
\be \label{c_sol}
c^2 \approx \frac{{\cal V}_{\hat{n} \hat{\tau}}^2}{9 ({\cal V}_{\hat{n} \hat{\tau}}^2 + {\cal V}_{\hat{n} \hat{n}}^2)} \,\,\, .
\ee
Note that no approximation is needed in order to obtain (\ref{c_sol}) from (\ref{VTN-Vntau_V_c}); the approximate sign $``\!\approx\!"$ in (\ref{c_sol}) is due solely to the $``\!\approx\!"$ signs in (\ref{VTN-Vntau_V_c}). Finally, substituting (\ref{c_sol}) back,in either of the two equations in (\ref{VTN-Vntau_V_c}), leads to the same result (just as in \cite{LA}):
\be \label{GenCC}
3 \,V \,{\cal V}_{\hat{n} \hat{n}} \,({\cal V}^2_{\hat{n} \hat{n}}+{\cal V}^2_{\hat{n} \hat{\tau}}) \,\approx \,{\cal V}^2_{\hat{n} \hat{\tau}} \,({\cal V}_{\hat{n} \hat{n}} {\cal V}_{\hat{\tau} \hat{\tau}}-{\cal V}^2_{\hat{n} \hat{\tau}}) \,\,\, .
\ee
This is the generalized consistency condition for slow-roll and rapid-turn inflation in the presence of the Gauss-Bonnet term. We underline that it is valid in the most general rapid-turn regime, defined by (\ref{RTcw}). Note that $V$ in (\ref{GenCC}) is the original potential, as it appears in (\ref{Action}), and not an effective expression of the form $(V - 3 f H^4)\approx (V- \frac{1}{3} f V^2)$\,, despite the form of the modified ``Hessian" components. Taking $f(\phi^I)=const$ in (\ref{GenCC}) gives precisely the condition found in \cite{LA,WIA} in the absence of the Gauss-Bonnet term.

The condition (\ref{GenCC}), together with (\ref{n_hat_SR}), relates the field-space metric $G_{IJ} (\phi^I)$, the scalar potential $V (\phi^I)$ and the coupling function $f (\phi^I)$\,. Since slow-roll and rapid-turn inflationary trajectories have to solve it, it can be used to numerically scan large sets of models in the present context, in order to search for such trajectories, as described in \cite{WIA} in the absence of the Gauss-Bonnet term.

\section{Dark Energy equation of state} \label{DEw}
\setcounter{equation}{0}

Until now, we have always assumed the slow-roll approximation, which is relevant for inflation in the early Universe. In this section, we turn to multifield models of Dark Energy (DE) in the late Universe, in which case slow-roll is not necessary. Specifically, we will study the equation of state for DE models within the current framework, which includes coupling the DE scalars $\phi^I$ to the Gauss-Bonnet (GB) term. Our main goal will be to investigate whether, in the presence of the GB term, a phantom regime is possible. This is of great interest in view of recent observations \cite{DESI,CCSGVM}. Note that, since there are no observational constraints on the DE slow-roll parameters, in the following we will not make any assumptions about the magnitudes of the quantities $\varepsilon$, $\delta_1$ and $\delta_2$ that were introduced in the previous Sections.

An important distinction between inflation and dark energy is that in the late Universe one cannot neglect the contribution of Dark Matter (DM). Including the latter modifies the equations of motion (\ref{gEoM}) and (\ref{gEoM3}) respectively to: 
\bea \label{EoMsDM}
3 H^2 &=& \left( \frac{1}{2} G_{IJ} \dot{\phi}_0^I \dot{\phi}_0^J + V - 3 H^2 \,\delta_1 \right) + \rho_{DM} \quad , \nn \\
- (2 \dot{H} + 3 H^2) &=& \left(\frac{1}{2} G_{IJ} \dot{\phi}_0^I \dot{\phi}_0^J - V + H^2 \,\delta_1 \delta_2 + 2 H^2 \,\delta_1 (1 - \varepsilon )\right)+ P_{DM} \quad ,
\eea
where $\rho_{DM}$ and $P_{DM}$ are the DM energy and pressure densities. Usually the dark matter pressure is assumed to vanish. Here we leave it in the expressions to be more general.  In (\ref{EoMsDM}) we have used the definitions (\ref{Param_def})-(\ref{SR_GB}) for future convenience. We will see below that the presence of dark matter is crucial, albeit indirectly, in order to obtain a phantom period. This is different from models that achieve a phantom-like behavior by including direct interactions between DE and DM in their actions \cite{BOVW,AKW}.

Let us now turn to computing the equation-of-state parameter $w_{DE}$ for a model of dark energy described by the action (\ref{Action}). The energy momentum tensor of the DE scalars $\phi^I$ is (see, for instance, \cite{HSSZ,JB}):
\be
T_{\mu \nu} = G_{IJ} \pd_{\mu} \phi^I \pd_{\nu} \phi^J - \frac{1}{2} g_{\mu \nu} G_{IJ} \pd_{\rho} \phi^I \pd^{\rho} \phi^J - g_{\mu \nu} V (\phi^I) - P_{\mu \alpha \nu \beta } \nabla^{\alpha} \nabla^{\beta} f (\phi^I) \,\,\, ,
\ee
where
\be \label{TmnP}
P_{\mu \alpha \nu \beta} = R_{\mu \alpha \nu \beta} + 2 g_{\mu [ \beta} R_{\nu ] \alpha} + 2 g_{\alpha [ \nu} R_{\beta ] \mu} + g_{\mu [ \nu} g_{\beta ] \alpha} R \,\,\, .
\ee
The corresponding energy and pressure densities are respectively:
\be
\rho = T_{00} \qquad , \qquad p = \frac{1}{3} \,g^{ij} T_{ij} \quad ,
\ee
where $i,j = 1,2,3$\,. Hence the equation of state parameter is given by:
\be \label{wDE}
w_{DE} = \frac{1}{3} \frac{(g^{ij} T_{ij})}{T_{00}} = \frac{\frac{1}{2} G_{IJ} \dot{\phi}_0^I \dot{\phi}_0^J - V - \frac{1}{3} g^{ij} P_{i \alpha j \beta} (\nabla^{\alpha} \nabla^{\beta} f)|_{\phi=\phi_0}}{\frac{1}{2} G_{IJ} \dot{\phi}_0^I \dot{\phi}_0^J + V - P_{0 \alpha 0 \beta} (\nabla^{\alpha} \nabla^{\beta} f)|_{\phi=\phi_0}} \,\,\, .
\ee 

In order to evaluate on the background $\phi^I = \phi_0^I (t)$ the terms with $\nabla^{\alpha} \nabla^{\beta} f $ in (\ref{wDE}), note that: 
\be
\nabla_{\alpha} \nabla_{\beta} f = \pd_{\alpha} (f_I \pd_{\beta} \phi^I ) - \Gamma_{\alpha \beta}^{\gamma} f_I \pd_{\gamma} \phi^I = f_{IJ} \pd_{\alpha} \phi^I \pd_{\beta} \phi^J + f_I \pd_{\alpha} \pd_{\beta} \phi^I - f_I \Gamma_{\alpha \beta}^{\gamma} \pd_{\gamma} \phi^I \,\,\, .
\ee
Now, using that the only non-vanishing Christoffel symbols for the background spacetime metric (\ref{Bkgr}) are:
\be
\Gamma^0_{ij} = a^2 H \delta_{ij} \qquad {\rm and} \qquad \Gamma^i_{j0} = H \delta^i_j \quad ,
\ee
we find:
\bea
\nabla_0 \nabla_0 f|_{\phi=\phi_0} &=& f_{IJ} \dot{\phi}_0^I \dot{\phi}_0^J + f_I \ddot{\phi}_0^I = \ddot{f} \quad , \nn \\
\nabla_0 \nabla_i f|_{\phi=\phi_0} &=& \nabla_i \nabla_0 f|_{\phi=\phi_0} = 0 \quad , \nn \\
\nabla_i \nabla_j f|_{\phi=\phi_0} &=& - a^2 H \,\delta_{ij} \,f_I \dot{\phi}_0^I \quad .
\eea
Using these results, we compute:
\bea \label{P0a0b}
P_{0 \alpha 0 \beta} (\nabla^{\alpha} \nabla^{\beta} f)|_{\phi=\phi_0} &=& [ P_{0000} (g^{00})^2 \,\nabla_0 \nabla_0 f + P_{0i0j} g^{ik} g^{jl} \nabla_k \nabla_l f ]|_{\phi=\phi_0} \nn \\
&=& - H f_I \,\dot{\phi}_0^I \,P_{0i0j} \,g^{ij} \quad ,
\eea
where we used that $R_{0000} = 0$ for the spacetime metric (\ref{Bkgr}). Similarly:
\be \label{Piaib}
P_{i \alpha j \beta} (\nabla^{\alpha} \nabla^{\beta} f)|_{\phi=\phi_0} = P_{i0j0} \,(g^{00})^2 \,\nabla_0 \nabla_0 f|_{\phi=\phi_0} - H f_I \,\dot{\phi}_0^I \,P_{ikjl} \,g^{kl} \quad .
\ee

Note that for the metric (\ref{Bkgr}), we have the following nonvnishing components of the Riemann tensor:
\be
R_{i0j0} = - a^2 (H^2 + \dot{H}) \delta_{ij} \quad , \quad R_{ijkl} = a^4 H^2 (\delta_{ik} \delta_{lj} - \delta_{il} \delta_{kj} ) \quad ,
\ee
while for the Ricci tensor and scalar curvature:
\be
R_{00} = - 3 (\dot{H}^2 + \dot{H}) \quad , \quad R_{ij} = a^2 (3H^2 + \dot{H}) \delta_{ij} \quad , \quad R = 6 (2 H^2 + \dot{H}) \quad .
\ee
Substituting those in (\ref{TmnP}) gives:
\be
P_{0i0j} \,g^{ij} = - 3 H^2 \quad , \quad P_{ikjl} \,g^{ij} g^{kl} = 6 (H^2 + \dot{H}) \quad .
\ee
Thus, (\ref{P0a0b}) and (\ref{Piaib}) imply respectively:
\be \label{P0a0b_f}
P_{0 \alpha 0 \beta} (\nabla^{\alpha} \nabla^{\beta} f)|_{\phi=\phi_0} = 3 H^3 f_I \dot{\phi}_0^I = 3 H^3 \dot{f}
\ee
and
\be \label{Piaib_f}
g^{ij} P_{i \alpha j \beta} (\nabla^{\alpha} \nabla^{\beta} f)|_{\phi=\phi_0} = - 3 H^2 \ddot{f} - 6 H f_I \dot{\phi}_0^I (H^2 + \dot{H}) = - 3 H^2 \ddot{f} - 6 H \dot{f} (H^2 + \dot{H}) \,\, .
\ee

Finally, substituting (\ref{P0a0b_f})-(\ref{Piaib_f}) in (\ref{wDE}), we obtain:
\be
w_{DE} = \frac{\frac{1}{2} G_{IJ} \dot{\phi}_0^I \dot{\phi}_0^J - V + H^2 \ddot{f} + 2 H \dot{f} (H^2 + \dot{H})}{\frac{1}{2} G_{IJ} \dot{\phi}_0^I \dot{\phi}_0^J + V - 3 H^3 \dot{f}} \,\,\, .
\ee
In terms of the slow-roll parameters introduced in (\ref{Param_def})-(\ref{SR_GB}), this can be rewritten as:
\be \label{wDE_r}
w_{DE} = \frac{\frac{1}{2} G_{IJ} \dot{\phi}_0^I \dot{\phi}_0^J - V + H^2 \,\delta_1 \delta_2 + 2 H^2 \,\delta_1 (1 - \varepsilon )}{\frac{1}{2} G_{IJ} \dot{\phi}_0^I \dot{\phi}_0^J + V - 3 H^2 \,\delta_1} \,\,\, .
\ee
Requiring $w_{DE} < -1$ implies:
\be \label{wDE_ph_ineq}
G_{IJ} \dot{\phi}_0^I \dot{\phi}_0^J < H^2 \delta_1 ( 1 + 2 \varepsilon - \delta_2) \quad .
\ee
Since $G_{IJ} \dot{\phi}_0^I \dot{\phi}_0^J >0$\,, this means that to have a phantom regime, one needs either:
\be \label{Op_ab}
a) \quad \delta_1>0 \quad {\rm and} \quad \delta_2 < 1 + 2 \varepsilon \qquad {\rm or} \qquad b) \quad \delta_1<0 \quad {\rm and} \quad \delta_2 > 1 + 2 \varepsilon \quad .
\ee
Recall that in the DE context there are no observational constraints on the magnitudes of the parameters $\varepsilon$\,, $\delta_1$ and $\delta_2$\,. (Except, of course, we always have $\varepsilon > 0$ in an expanding Universe.) One should also keep in mind that these parameters are functions of time. Hence options $a)$ and $b)$ above are not discrete choices. There may be solutions such that, during their evolution a transition between $a)$ and $b)$ occurs. This could be particularly relevant for the existence of a {\it transient} phantom-like regime, since for $\delta_1 >0$ the coupling $f (\phi)$ increases, while for $\delta_1< 0 $ it decreases with time. 

So far, dark matter has not played a role in the considerations of this Section. However, notice that, due to the equations of motion (\ref{EoMsDM}), the DE equation of state (\ref{wDE_r}) can be written as:
\be
w_{DE} = \frac{- (2 \dot{H} + 3 H^2) - P_{DM}}{3H^2 - \rho_{DM}} = \frac{- 3H^2 (1 - \frac{2}{3} \varepsilon) - P_{DM}}{3H^2 - \rho_{DM}} \quad .
\ee
Hence, in the absence of dark matter, this would become $w_{DE} = -1 + \frac{2}{3} \varepsilon$\,, which always satisfies $w_{DE} > -1$ due to $\varepsilon > 0$\,. So the presence of dark matter is crucial for the possibility of a phantom regime, even though in our case there is no direct DE-DM interaction in the Lagrangian. As a side note, the discussion in this paragraph means that during inflation, when dark matter is negligible, phantom-like behavior is not possible in any model within this framework.

Of course, the result (\ref{wDE_r}) is consistent with the field equations. However, the derivation here, starting from the energy-momentum tensor, may be instructive for generalizations to other higher-curvature terms. We considered the Gauss-Bonnet term because it provides a simple first step. Namely, since it is a topological invariant in four dimensions, its only contribution to the dynamics comes from derivatives of the coupling function $f(\phi^I)$. However, there are many other higher-curvature terms that are of physical interest in the literature, for instance, in the context of string theory \cite{GrSch,GrWi,GVdVZ,GV,LMSS,LMSS2} or in cosmological higher-curvature gravities \cite{MM,AEJ,CGO}. The latter are an interesting class of four-dimensional gravitational theories, which have second-order equations of motion on a cosmological background, even though they are not embedded in a known UV completion. Although other higher-curvature terms than the GB one will lead to quantitatively different results, qualitatively the relevant computations would be similar to the one we performed here. Thus, it is natural to expect that such terms will also enable a phantom regime for suitable ranges of the relevant parameters. It would be very interesting to study this topic systematically, and we hope to return to it in the future.

\section{Discussion} \label{Disc}
\setcounter{equation}{0}

In this paper, we initiated the study of the implications of higher-curvature terms for the rapid-turn regime of multifield cosmology. Considering two-field models coupled to the Gauss-Bonnet (GB) term, we derived the GB-modified relation between the inflationary slow-roll parameters $\varepsilon$ and $\varepsilon_V$\,, as well as the GB contributions to the Hessian components of the scalar potential $V$ that determine $\eta_V$\,. Notably, we found that $\varepsilon$ and $\varepsilon_V$ can differ appreciably even for non-large turning rates, due to the GB coupling; and similarly for $\eta_{\parallel}$ and $\eta_V$\,. These results show that including the higher-curvature term broadens the possibilities for realizing the rapid-turn regime and for achieving consistency with the swampland conjectures. We also derived a necessary condition that two-field slow-roll and rapid-turn inflationary trajectories must satisfy. It relates the field-space metric, the scalar potential, and the GB-coupling function. This relation can be used to perform numerical scans over large sets of models in the search for such trajectories, a la \cite{WIA}. Finally, we discussed how the GB term enriches multifield dark-energy models by enabling phantom-like behavior. Although we did not include a direct non-gravitational interaction between dark energy and dark matter, we found that a non-negligible amount of dark matter is crucial for the occurrence of a phantom period.

It would be very interesting to investigate how the GB contributions can affect rapid-turn models of primordial black hole generation \cite{PSZ,FRPRW,LA1_pbh,LA2_pbh} or dark energy clustering \cite{ASSV,EASV,ADGW2,AA}. It is also worth considering other higher-curvature terms in the context of both inflation and dark energy. Recall that standard single-scalar or multi-field DE cosmological models do not allow a phantom regime. Thus, it would be especially interesting to explore which (kinds of) higher-curvature terms, and under what conditions, can accommodate phantom-like behavior. Another topic worth investigating is whether including the GB term, or other higher-curvature terms, can facilitate embedding rapid-turn inflation in string compactifications. Although there are many field-theoretic inflationary models in the rapid-turn regime, stringy realizations have not yet been found. As an initial step in that direction, one may address embedding rapid-turn inflation in supergravity. This has proven quite difficult, as discussed in \cite{ACPRZ,Freigang:2023ogu}. An additional ingredient, such as a higher-curvature term, could be very helpful in that regard. We hope to investigate this topic in the future.

\section*{Acknowledgements}

\noindent We would like to thank D. Andriot, S. Parameswaran, M. Rocek, A. Tomasiello, C. Vafa and I. Zavala for useful discussions. We also thank the Stony Brook Simons Physcs Summer Workshop for hospitality during the completion of this work. L.A. has received partial support from the Bulgarian NSF grant KP-06-N68/3.

\end{document}